\documentclass[conference, a4paper]{IEEEtran}
\IEEEoverridecommandlockouts
\usepackage[T1]{fontenc}

\usepackage{cite}
\usepackage{xurl} 
\usepackage[colorlinks=false, pdfborder={0 0 0}]{hyperref}

\usepackage{graphicx}

\usepackage{siunitx}
\usepackage{multirow}
\usepackage{amsmath,amssymb,amsfonts,mathtools}
\usepackage{placeins}
\usepackage{tabularx,multirow,multicol,booktabs,colortbl,array,makecell}
\usepackage{algorithm,algpseudocode}
\usepackage{enumitem}
\usepackage[capitalise]{cleveref}

\usepackage[nolist]{acronym}
\usepackage[english]{babel}
\usepackage{circledsteps} 
\usepackage{microtype} 

\makeatletter
\@ifundefined{AC@verridelabel}{%
  \newcommand*{\org@overridelabel}{}%
  \let\org@overridelabel\@verridelabel
  \@ifpackagelater{acronym}{2015/03/21}{
    \renewcommand*{\@verridelabel}[1]{%
      \@bsphack
      \protected@write\@auxout{}{\string\AC@undonewlabel{#1@cref}}%
      \org@overridelabel{#1}%
      \@esphack
    }%
  }{
    \renewcommand*{\@verridelabel}[1]{%
      \@bsphack
      \protected@write\@auxout{}{\string\undonewlabel{#1@cref}}%
      \org@overridelabel{#1}%
      \@esphack
    }%
  }%
}{%
  \let\org@AC@verridelabel\AC@verridelabel
  \renewcommand*{\AC@verridelabel}[1]{%
    \org@AC@verridelabel{#1}%
    \@bsphack
    \protected@write\@auxout{}{\string\AC@undonewlabel{#1@cref}}%
    \@esphack
  }%
}
\makeatother

\usepackage{tikz}
\usetikzlibrary{arrows.meta, positioning, calc}

\usepackage{pifont}
\usepackage{subcaption}

\usepackage{balance}
\usepackage{mleftright}

\usepackage{pgfplots}
\pgfplotsset{compat=1.18}
\usepgfplotslibrary{groupplots}
\pgfplotsset{ 
  /pgfplots/layers/custom/.define layer set={
    axis background,axis grid,axis ticks,main,axis lines,axis tick labels,
    axis descriptions,axis foreground
  }{/pgfplots/layers/standard},
}

\makeatletter
\newcommand\resetstackedplots{
  \makeatletter
  \pgfplots@stacked@isfirstplottrue
  \makeatother
  \addplot [forget plot, draw=none] coordinates{(1,0) (2,0) (3,0)};
}
\makeatother

\DeclareMathOperator{\pred}{pred}

\definecolor{myblue}{HTML}{00549f}
\definecolor{myred}{HTML}{a11035}
\definecolor{mypurple}{HTML}{7a6fac}
\definecolor{myorange}{HTML}{f6a800}
\definecolor{mygreen}{HTML}{57ab27}
\definecolor{myviolett}{HTML}{612158}

\usepackage[some]{background}
\usepackage{stackengine}
\setstackEOL{\\}
\setstackgap{L}{\normalbaselineskip}
\SetBgContents{\color{gray}{\tiny \Longstack{
      PREPRINT - Accepted for publication at the 34th IFIP/IEEE International Conference on Very Large Scale Integration SoC (VLSI-SoC),
      \\October \numrange[range-phrase = --]{11}{14}, 2026, in Limassol, Cyprus.\\
      \textcopyright 2026 IEEE. Personal use of this material is permitted. Permission from IEEE must be obtained for all other uses, in any current or future media,\\
      including reprinting/republishing this material for advertising or promotional purposes, creating new collective works, for resale\\
      or redistribution to servers or lists, or reuse of any copyrighted component of this work in other works.
}}}
\SetBgPosition{current page.north} 
\SetBgVshift{-0.52cm}               
\SetBgOpacity{1.0}
\SetBgAngle{0}
\SetBgScale{1.8}

\def\BibTeX{{\rm B\kern-.05em{\sc i\kern-.025em b}\kern-.08em
T\kern-.1667em\lower.7ex\hbox{E}\kern-.125emX}}
\begin{document}
\BgThispage
\renewcommand{\figurename}{Fig.} 
\bstctlcite{IEEEexample:BSTcontrol}

\begin{acronym}
  \acro{1t1r}[1T1R]{One-Transistor One-Resistor}
  \acro{adc}[ADC]{Analog-to-Digital Converter}
  \acro{bip}[BIP]{Binary Integer Programming}
  \acro{cim}[CIM]{Computing-in-Memory}
  \acro{cnn}[CNN]{Convolutional Neural Network}
  \acro{cpu}[CPU]{Central Processing Unit}
  \acro{dac}[DAC]{Digital-to-Analog Converter}
  \acro{dag}[DAG]{Directed Acyclic Graph}
  \acro{dse}[DSE]{Design Space Exploration}
  \acro{fpga}[FPGA]{Field-Programmable Gate Array}
  \acro{gemm}[GeMM]{General Matrix Multiplication}
  \acro{gpeu}[GPEU]{General-Purpose Execution Unit}
  \acro{gpu}[GPU]{Graphics Processing Unit}
  \acro{ilp}[ILP]{Integer Linear Programming}
  \acro{isa}[ISA]{Instruction Set Architecture}
  \acro{lp}[LP]{Linear Programming}
  \acro{mac}[MAC]{Multiply-Accumulate}
  \acro{ml}[ML]{Machine Learning}
  \acro{mip}[MIP]{Mixed-Integer Programming}
  \acro{mvm}[MVM]{Matrix-Vector Multiplication}
  \acro{mvmu}[MVMU]{Matrix-Vector Multiplication Unit}
  \acro{nn}[NN]{Neural Network}
  \acro{onnx}[ONNX]{Open Neural Network Exchange}
  \acro{ram}[RAM]{Random Access Memory}
  \acro{relu}[ReLU]{Rectified Linear Unit}
  \acro{resnet}[ResNet]{Residual Neural Network}
  \acro{rram}[RRAM]{Resistive Random Access Memory}
  \acro{soc}[SoC]{System-on-a-Chip}
  \acro{sram}[SRAM]{Static Random Access Memory}
  \acro{yolo}[YOLO]{You Only Look Once}
\end{acronym}

\title{
  Optimizing ML Workload Partitioning between CPUs and CIM Accelerators for Heterogeneous Computing
}

\def\finalpaper{1}

\if\finalpaper1
{
  \author{
    \IEEEauthorblockN{Joel Klein, 
      Rebecca Pelke, 
      Roberto Laudani, 
      Jan Moritz Joseph, 
    Rainer Leupers}
    \IEEEauthorblockA{
      Institute for Communication Technologies and Embedded Systems,
      RWTH Aachen University, Germany\\
      \{klein, pelke, laudani, joseph, leupers\}@ice.rwth-aachen.de
    }
    \thanks{This work was funded by the Federal Ministry of Research, Technology and Space, Germany, in the project NeuroSys II (03ZU2106CA).}
  }
}
\else
\author{
  \IEEEauthorblockN{Authors are removed for submission}
  \\
  \\
  \vspace{-0.5cm}
  \IEEEauthorblockA{Affiliations are removed for submission}
  \thanks{Funding agencies are removed for submission}
  \\
}
\fi

\maketitle

\begin{abstract}
  \Ac{cim} accelerators execute \acp{mvm} in memory, making them a compelling solution for \ac{ml} workloads.
  However, existing \ac{ml} workload partitioning approaches for \ac{cim} accelerators do not fully account for \ac{rram} constraints such as limited memory, high write latency, and limited endurance.
  They also neglect parallelism, low-level architectural effects, or the \ac{cpu} as a complementary compute resource.

  To address these limitations, we propose an \ac{ilp}-based workload partitioning framework for heterogeneous \ac{cpu}--\ac{cim} systems.
  It minimizes end-to-end inference latency under \ac{rram} constraints, captures parallelism, and combines empirical profiling with analytical models.
  Using our framework, heterogeneous \ac{cpu}--\ac{cim} execution achieves speedups of up to \(\qty{30.9}{\times}\) over \ac{cpu}-only execution on an edge \ac{cpu} and \(\qty{7.3}{\times}\) over a high-performance \ac{cpu}.
  A \ac{dse} yields further design insights for future \ac{cim} accelerators.
\end{abstract}

\acresetall 

\section{Introduction}
\label{sec:introduction}

Efficient execution of \ac{ml} workloads requires new hardware architectures. \Ac{cim} fuses computation and memory to address the von Neumann bottleneck~\cite{amrouch2021ReliableInMemoryComputing}, executing \acp{mvm} with \(\mathcal{O}\mleft(1\mright)\) latency using crossbar arrays~\cite{shafiee2016ISAACConvolutionalNeural,chi2016PRIMENovelProcessinginMemory,ankit2019PUMAProgrammableUltraefficient}. \Ac{rram} is a promising \ac{cim} candidate due to its high device density, low power consumption, non-volatility, and CMOS \mbox{compatibility~\cite{wan2022ComputeinmemoryChipBased,zahoor2020ResistiveRandomAccess,amrouch2021ReliableInMemoryComputing}}. However, \ac{cim} accelerators rely on a host \ac{cpu} to execute unsupported operators, requiring workload partitioning between \ac{cpu} and \ac{cim} accelerator~\cite{ankit2019PUMAProgrammableUltraefficient}.

Furthermore, \ac{cim} storage is limited by the available crossbars~\cite{gao2024StaticSchedulingWeight}. Modern \acp{cnn} often exceed this capacity~\cite{he2016DeepResidualLearning,sandler2018MobileNetV2InvertedResiduals,jocher2020UltralyticsYOLOv5,park2025COMPASSCompilerFramework,gao2024StaticSchedulingWeight}. Moreover, \ac{rram} write latency and endurance limits make dynamic weight replacement impractical~\cite{zahoor2020ResistiveRandomAccess,swaidan2019RRAMEnduranceRetention,wan2022ComputeinmemoryChipBased}. Hence, deployment requires static assignment of selected operators to the \ac{cim} accelerator.

Existing \ac{ml} partitioning approaches~\cite{kang2017NeurosurgeonCollaborativeIntelligence,viramontes2023NeuralNetworkPartitioning,zeng2021CoEdgeCooperativeDNN,taufique2025HiDPHierarchicalDNN} treat devices as black boxes, neglect low-level architectural characteristics or assume sequential execution. \Ac{cim}-specific methods either assume full-model fit~\cite{li2023MathematicalFrameworkOptimizing,qu2024CIMMLCMultilevelCompilation} or rely on weight reprogramming~\cite{gao2024StaticSchedulingWeight,park2025COMPASSCompilerFramework}. Furthermore, workload partitioning between the host \ac{cpu} and the \ac{cim} accelerator remains an open problem.

\begin{figure}[tb]
  \centering
  \includegraphics[width=\linewidth]{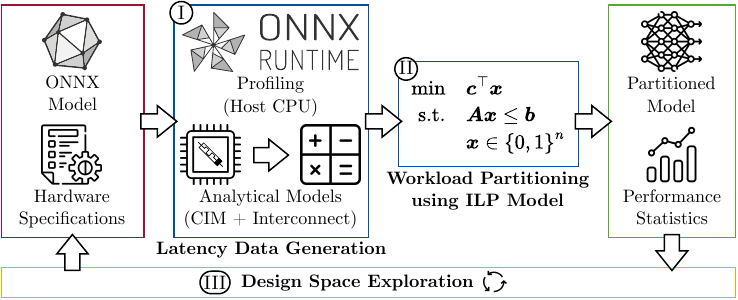}
  \caption{Overview of the proposed partitioning framework}
  \label{fig:workload_partitioning_tool}
\end{figure}

We propose a system-level workload partitioning framework for heterogeneous \ac{cpu}--\ac{cim} systems. It covers latency characterization and static operator assignment. The resulting partitioning between \ac{cpu} and \ac{cim} achieves up to \(\qty{30.9}{\times}\) and \(\qty{7.3}{\times}\) speedup over \ac{cpu}-only execution on ARM edge and x86 \acp{cpu}, respectively. \cref{fig:workload_partitioning_tool} illustrates the framework's components. We highlight the following contributions:

\begin{enumerate}[label=\Circled{\Roman*}]
  \item A \textbf{hybrid latency characterization} methodology, integrating empirical host \ac{cpu} profiling with an analytical \ac{cim} accelerator and interconnect performance model.
  \item An \textbf{\ac{ilp}} formulation for static operator assignment that minimizes the end-to-end inference latency accounting for limited \ac{cim} memory budget, operator support, and parallelism constraints.
  \item A \textbf{\ac{dse}} using the proposed framework to analyze the impact of model architectures and hardware parameters on system performance, providing design insights for future \ac{cim} accelerators.
\end{enumerate}

\cref{sec:background,sec:related_work} provide background and review prior work. \cref{sec:partitioning_framework} presents our framework. \cref{sec:results} reports results and a \ac{dse}. \cref{sec:conclusion} concludes the paper.
\section{Background}
\label{sec:background}

This section presents background related to \ac{rram}-based \ac{cim} accelerators and the \ac{ilp} concepts used in this work.

\subsection{\acs{rram}-based \acs{cim} Accelerators}
\label{subsec:rram_based_cim_accelerators}

\Ac{rram} is a memristive non-volatile memory technology that stores data as resistance states~\cite{zahoor2020ResistiveRandomAccess,wan2022ComputeinmemoryChipBased,amrouch2021ReliableInMemoryComputing}. By arranging \ac{rram} cells in crossbar arrays, \ac{cim} accelerators perform analog \acp{mvm} via Ohm's and Kirchhoff's laws~\cite{shafiee2016ISAACConvolutionalNeural,chi2016PRIMENovelProcessinginMemory,ankit2019PUMAProgrammableUltraefficient}. Since programming \ac{rram} cells is slow and repeated writes degrade device endurance~\cite{zahoor2020ResistiveRandomAccess,swaidan2019RRAMEnduranceRetention}, a weight-stationary dataflow, where weights are programmed once before inference, is particularly well-suited for \ac{rram}-based \ac{cim} accelerators~\cite{shafiee2016ISAACConvolutionalNeural,ankit2019PUMAProgrammableUltraefficient}.

The PUMA architecture~\cite{ankit2019PUMAProgrammableUltraefficient} is hierarchically organized into \ac{cim} cores, connected by an internal bus.
Each core contains a control unit, instruction memory, a \ac{mvmu}, a \ac{gpeu}, and a data buffer.
The \ac{mvmu} comprises multiple \ac{rram} crossbars with peripheral circuits.
The \ac{gpeu} performs simple arithmetic and a limited set of functions, including activation functions.

\subsection{\acl{ilp}}
\label{subsec:ilp}

\Ac{ilp} is a class of optimization problems where decision variables are restricted to integers. The objective is to maximize or minimize a linear function with linear constraints. \Ac{lp} allows continuous variables. \Ac{mip} mixes integer and continuous variables.

State-of-the-art solvers~\cite{gurobioptimizationllc2026GurobiOptimizerReference} combine fundamental methods such as branch-and-bound with heuristics to identify optimal or near-optimal solutions. The relative \ac{mip} gap measures solution quality as \(\frac{z_{\mathrm{inc}} - z_{\mathrm{bound}}}{\mleft|z_{\mathrm{inc}}\mright|}\), where \(z_{\mathrm{inc}}\), the incumbent, is the best known feasible integer solution and \(z_{\mathrm{bound}}\) is the proven bound obtained from the \ac{lp} relaxation. However, a large gap does not necessarily indicate a suboptimal solution, since the \ac{lp} bound may be significantly looser than the true \ac{ilp} optimum.

\section{Related Work}
\label{sec:related_work}

Prior research has explored \ac{ml} workload partitioning for heterogeneous systems and \ac{cim} accelerators specifically.

Neurosurgeon~\cite{kang2017NeurosurgeonCollaborativeIntelligence} splits \ac{ml} execution between a mobile device and a datacenter at a layer granularity, selecting the optimal split point via per-layer prediction models.
Viramontes et al.~\cite{viramontes2023NeuralNetworkPartitioning} extend this to a multi-device \ac{ilp} formulation for edge-hub-cloud hierarchies, using weight preloading and multiple device transitions to minimize inference latency. CoEdge~\cite{zeng2021CoEdgeCooperativeDNN} and HiDP~\cite{taufique2025HiDPHierarchicalDNN} further generalize to cooperative and hierarchical partitioning across heterogeneous edge nodes.
These methods largely treat devices as black boxes, rely on high-level latency estimates, or assume sequential layer execution.

Regarding \ac{cim}-specific approaches, Li et al.~\cite{li2023MathematicalFrameworkOptimizing} optimize crossbar allocation by determining the per-layer weight replication factor for static mapping, and CIM-MLC~\cite{qu2024CIMMLCMultilevelCompilation} provides a multi-level compilation stack for operator scheduling across device, circuit, and architecture tiers.
Both assume the complete model fits on the accelerator.
Gao et al.~\cite{gao2024StaticSchedulingWeight} address the resource-constrained case by statically scheduling weight programming to minimize reprogramming overhead.
COMPASS~\cite{park2025COMPASSCompilerFramework} addresses capacity limitations through weight partitioning, replication, and dynamic weight replacement.
For \ac{rram}-based targets, however, this degrades device endurance and introduces write latency, making reliable long-term deployment infeasible.
Furthermore, none of the prior works specifically addresses joint \ac{cpu}--\ac{cim} execution of \ac{ml} workloads.

In contrast, we target system-level \ac{cpu}--\ac{cim} partitioning, combining empirical \ac{cpu} profiling with analytical models. Our \ac{ilp} formulation accounts for memory constraints, operator support, and parallelism, with a weight-stationary dataflow.
\section{Partitioning Framework}
\label{sec:partitioning_framework}

\begin{figure}[!t]
  \centering
  \includegraphics[width=\linewidth]{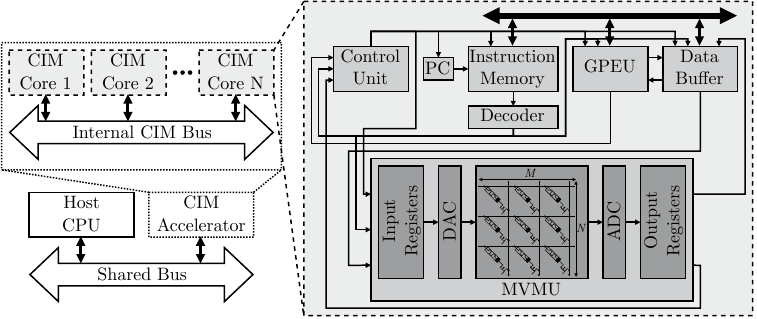}
  \caption{Overview of the heterogeneous hardware setup}
  \label{fig:hardware_setup}
\end{figure}

The framework targets a hardware as illustrated in \cref{fig:hardware_setup}. The platform comprises a single-core host \ac{cpu} and a PUMA-based \ac{cim} accelerator~\cite{ankit2019PUMAProgrammableUltraefficient} connected through a shared bus. The \ac{cim} accelerator comprises multiple \ac{cim} cores (see \cref{subsec:rram_based_cim_accelerators}).

\subsection{Hybrid Latency Characterization}
\label{subsec:hybrid_latency}

To accurately characterize operator execution latencies on both devices, we adopt a hybrid approach combining empirical \ac{cpu} profiling with analytical \ac{cim} and interconnect modeling.

\subsubsection{Host \acs{cpu} Profiling}
\label{subsubsec:host_cpu_profiling}

The operator latency on the \ac{cpu}, \(T_{\mathrm{host},v}\), is measured by profiling each operator using \ac{onnx} Runtime~\cite{onnxdevelopers2026ONNXOpenNeural,onnxruntimedevelopers2021ONNXRuntime} on the target hardware. This empirical approach captures all architectural and optimization effects, which are difficult to model analytically, while supporting flexible replacement of the host \ac{cpu}. We perform multiple iterations with warmup runs to eliminate startup effects, using the median latency for robustness. For model-level statistics, we enable all available graph and layout optimizations.

\subsubsection{CIM Accelerator Modeling}
\label{subsubsec:cim_accelerator_modeling}

We model the \ac{cim} accelerator latency analytically, which enables evaluation across a broad range of accelerator configurations. \Ac{rram} crossbars execute one \ac{mvm} per input vector in \(\mathcal{O}\mleft(1\mright)\) time~\cite{shafiee2016ISAACConvolutionalNeural,chi2016PRIMENovelProcessinginMemory,ankit2019PUMAProgrammableUltraefficient}. However, limited crossbar size requires large operators to span multiple crossbars~\cite{shafiee2016ISAACConvolutionalNeural,pelke2023MappingCNNsMulticore,pelke2026MixedPrecisionTrainingCompilation,li2023MathematicalFrameworkOptimizing}. The number of crossbars \(N_{\mathrm{crossbar},v}\) required for an operator \(v\) is defined as:
\begin{equation}
  N_{\mathrm{crossbar},v} = \mleft\lceil \frac{M_v}{M} \mright\rceil \cdot \mleft\lceil \frac{N_v}{N} \mright\rceil,
  \label{eq:number_of_crossbars}
\end{equation}
where \(M_v\) and \(N_v\) are the dimensions of the weight matrix for operator \(v\), and \(M\) and \(N\) are the dimensions of a single \ac{rram} crossbar, as illustrated in \cref{fig:hardware_setup}.

Operators without a 2D weight matrix are converted to a \ac{gemm}. For convolutions, this is achieved using \textit{im2col}~\cite{jeon2025LowRankCompressionIMC,pelke2023MappingCNNsMulticore}. The number of \acp{mvm} \(N_{\mathrm{mvm},v}\) required to execute a \ac{gemm} equals the number of input vectors applied to the crossbars. For instance, a Conv2D layer \(\mleft(C_{\mathrm{in}}, H_{\mathrm{in}}, W_{\mathrm{in}}\mright) \rightarrow \mleft(C_{\mathrm{out}}, H_{\mathrm{out}}, W_{\mathrm{out}}\mright)\) with kernel size \(\mleft(H_{\mathrm{K}},W_{\mathrm{K}}\mright)\) can be expressed as \(N_{\mathrm{mvm},v} = H_{\mathrm{out}} \cdot W_{\mathrm{out}}\) \acp{mvm} with dimensions \(M_v = C_{\mathrm{out}}\) and \(N_v = C_{\mathrm{in}} \cdot H_{\mathrm{K}} \cdot W_{\mathrm{K}}\).

The operator latency on the accelerator, \(T_{\mathrm{acc},v}\), is modeled as the sum of the load, compute, and store latency~\cite{ankit2019PUMAProgrammableUltraefficient,pelke2023MappingCNNsMulticore}:
\begin{equation}
  T_{\mathrm{acc},v} = T_{\mathrm{load},v} + T_{\mathrm{mvm},v} + T_{\mathrm{store},v}.
\end{equation}
Here, \(T_{\mathrm{mvm},v} = N_{\mathrm{mvm},v} \cdot T_{\mathrm{mvm}}\), with \(T_{\mathrm{mvm}}\) denoting the latency of a single \ac{mvm}. The latency is independent of the number of crossbars, as these operate in parallel~\cite{ankit2019PUMAProgrammableUltraefficient,pelke2023MappingCNNsMulticore}.

Load and store latencies depend on the internal bandwidth \(B_\mathrm{cim}\), element size \(b_\mathrm{elem}\), and element counts \(N_{\mathrm{in},v}\) and \(N_{\mathrm{out},v}\):
\begin{equation}
  T_{\mathrm{load},v} = \frac{b_\mathrm{elem} \cdot N_{\mathrm{in},v}}{B_\mathrm{cim}}, \quad T_{\mathrm{store},v} = \frac{b_\mathrm{elem} \cdot N_{\mathrm{out},v}}{B_\mathrm{cim}}.
\end{equation}

We assume the partial-sum accumulation and activation application on the \ac{gpeu} are pipelined, thus not contributing to the overall latency~\cite{ankit2019PUMAProgrammableUltraefficient}.

\subsubsection{Inter-device Transfer Modeling}
\label{subsubsec:inter_device_transfer_modeling}

For directly data-de\-pen\-dent operators \(u\) and \(v\) mapped to different devices, the output of \(u\) is transferred over the shared bus. The transfer latency is:
\begin{equation}
  T_{\mathrm{transfer},\mleft(u,v\mright)} = \frac{b_\mathrm{elem} \cdot N_{\mathrm{out},u}}{B_\mathrm{shared}},
\end{equation}
where \(B_\mathrm{shared}\) is the bus bandwidth between \ac{cpu} and \ac{cim}.

\subsection{\acs{ilp}-based Workload Partitioning}
\label{subsec:ilp_partitioning}

We formulate the operator partitioning as an \ac{ilp} on a \ac{dag} \(G = (V, E)\), where \(V\) is the operator set and each directed edge \((u,v) \in E \subseteq V \times V\) encodes a direct data dependency from \(u\) to \(v\).

Each operator \(v\) has a binary decision variable \(x_v\), where \(x_v = 1\) maps to host \ac{cpu} and \(x_v = 0\) maps to \ac{cim}. Only a subset of operators \(V_{\mathrm{acc}} \subseteq V\) is supported by the accelerator:
\begin{equation}
  x_v \in \mleft\{0,1\mright\}, \;\: \forall v \in V_{\mathrm{acc}},
  \quad x_v = 1, \;\: \forall v \in V \setminus V_{\mathrm{acc}}.
\end{equation}

The objective of the \ac{ilp} is to minimize the makespan \({T \in \mathbb{R}_{\geq 0}}\), constrained by the finish times of all terminal nodes:
\begin{equation}
  T \geq f_v, \quad \forall v \in \mleft\{u \in V \mid \nexists\, w \in V \colon \mleft(u,w\mright) \in E\mright\}.
\end{equation}

The finish time \(f_v\) depends on the start time \(s_v \in \mathbb{R}_{\geq 0}\) and the device-specific computation latency:
\begin{equation}
  f_v = s_v + x_v \cdot T_{\mathrm{host},v} + (1 - x_v) \cdot T_{\mathrm{acc},v}, \quad \forall v \in V.
\end{equation}

The start time \(s_v\) is bounded by all predecessors \(\pred\mleft(v\mright) = \{u \in V \mid \mleft(u,v\mright) \in E\}\) and any required transfer latency:
\begin{equation}
  s_v \geq f_u + z_{u,v} \cdot T_{\mathrm{transfer},\mleft(u,v\mright)}, \quad \forall u \in \pred\mleft(v\mright),
\end{equation}

The binary auxiliary variable \(z_{u,v}\) is \num{1} when \(u\) and \(v\) are assigned to different devices, and \num{0} otherwise, enforced by:
\begin{equation}
  z_{u,v} \geq x_u - x_v, \quad z_{u,v} \geq x_v - x_u, \quad \forall \mleft(u,v\mright) \in E.
\end{equation}

\subsubsection{Parallel Execution}
\label{subsubsec:parallel_execution}

Operators without data dependencies can execute in parallel on separate crossbars or across the host and the \ac{cim} accelerator. We identify such pairs via the transitive closure \(G^{+} = (V, E^{+})\), where \(E^{+}\) contains all pairs \((u,v)\) connected by a path in \(G\). The parallelizable pairs are:
\begin{equation}
  P = \mleft\{\mleft\{u,v\mright\} \subseteq V \mid u \neq v, \mleft(u,v\mright) \notin E^{+}, \mleft(v,u\mright) \notin E^{+}\mright\}.
\end{equation}

To serialize pairs when both operators are assigned to the host, we introduce a binary ordering variable \(o_{u,v}\) for each \(\{u,v\} \in P\), where \(o_{u,v} = 1\) if \(u\) precedes \(v\):
\begin{alignat}{2}
  s_v &\geq f_u - M \cdot \mleft(3 - x_u - x_v - o_{u,v}\mright), &\quad& \forall \{u,v\} \in P, \\
  s_u &\geq f_v - M \cdot \mleft(2 - x_u - x_v + o_{u,v}\mright), &\quad& \forall \{u,v\} \in P.
\end{alignat}

We set \(M\) to an upper bound on the inference latency, where
\(T_{\mathrm{acc},v} \coloneqq 0\) for \(v \notin V_{\mathrm{acc}}\):
\begin{equation}
  M = \sum_{v \in V} \max\mleft(T_{\mathrm{host},v}, T_{\mathrm{acc},v}\mright) + {\sum_{(u,v) \in E}} T_{\mathrm{transfer},\mleft(u,v\mright)},
\end{equation}
which deactivates the serializing constraint if either operator is assigned to the \ac{cim} accelerator or the order does not apply.

\subsubsection{Resource Constraint}
\label{subsubsec:resource_constraint}

Finally, the available \ac{rram} crossbar count \(C \in \mathbb{N}\) limits how many operators can be offloaded:
\begin{equation}
  \sum_{v \in V_{\mathrm{acc}}} (1 - x_v) \cdot N_{\mathrm{crossbar},v} \leq C,
\end{equation}
where \(N_{\mathrm{crossbar},v}\) is the number of \ac{rram} crossbars required to execute operator \(v\) on the accelerator, as defined in \eqref{eq:number_of_crossbars}.

\section{Results}
\label{sec:results}

In this section, we evaluate our framework in terms of speedup, operator allocation, and \ac{ilp} scalability on representative \acp{cnn}. All \ac{ilp} instances are solved using Gurobi~\cite{gurobioptimizationllc2026GurobiOptimizerReference}.

Experiments are conducted on two single-core hosts: an AMD Ryzen 9 3900X (\qty{3.8}{\giga\hertz}, \qty{32}{\giga\byte} DDR4) for high-performance x86 and an ARM Cortex-A72 (\qty{1.5}{\giga\hertz}, \qty{4}{\giga\byte} LPDDR4) for edge deployment. The \ac{cpu}--\ac{cim} bus bandwidth is \qty[per-mode=symbol]{20.48}{\giga\byte\per\second} on x86 and \qty[per-mode=symbol]{10.24}{\giga\byte\per\second} on ARM. We sweep \numlist{25;50;100;200} crossbars and \qtylist{50;100;500;1000}{\nano\second} \ac{mvm} latency. This \ac{dse} covers both hardware parameters and model architecture effects on system performance. Each crossbar stores up to \numproduct[product-symbol = \times]{256 x 256} weights, and the internal \ac{cim} bus bandwidth matches the shared \ac{cpu}--\ac{cim} bus bandwidth.

We evaluate \acs{resnet}-18 and \acs{resnet}-50~\cite{he2016DeepResidualLearning} as convolution-heavy architectures, MobileNetV2~\cite{sandler2018MobileNetV2InvertedResiduals} for depthwise separable convolutions, and \acs{yolo}v5n~\cite{jocher2020UltralyticsYOLOv5} for complex branching patterns. All models are quantized to \qty{8}{\bit} integer precision~\cite{jacob2018QuantizationTrainingNeural}.

\subsection{\acs{cim} Acceleration Speedup}
\label{subsec:cim_speedup}

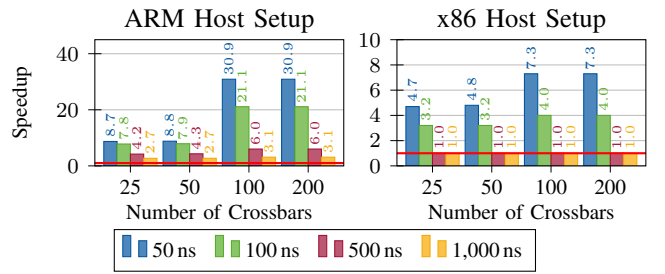
\begin{figure}[b]
  \centering
  \begin{tikzpicture}
  \begin{groupplot}[
      group style={
        group size=2 by 1,
        horizontal sep=2em,
        xlabels at=edge bottom,
        ylabels at=edge left,
      },
      ybar=0pt,
      width=0.55\linewidth,
      height=3.25cm,
      enlarge x limits=0.2,
      ylabel={Speedup},
      ylabel style={font=\footnotesize},
      xlabel={Number of Crossbars},
      xlabel style={font=\footnotesize, yshift=+3pt},
      symbolic x coords={25, 50, 100, 200},
      xtick=data,
      xticklabels={25, 50, 100, 200},
      tick label style={font=\footnotesize},
      x tick label style={yshift=+1pt},
      major tick length=5pt,
      tick align=center,
      tick pos=left,
      tick style={solid, black},
      ymajorgrids=true,
      major grid style={solid, gray!50},
      set layers=custom,
      /pgf/bar width=5pt,
      nodes near coords,
      nodes near coords align={vertical},
      every node near coord/.append style={
        font=\tiny,
        rotate=90,
        anchor=west,
        xshift=-2pt,
        /pgf/number format/.cd,
        fixed,
        precision=1,
        zerofill,
        /tikz/.cd,
      },
      legend style={
        at={(0.5,-0.45)},
        anchor=north,
        legend columns=-1,
        font=\footnotesize,
        draw=black,
        fill=white,
        /tikz/every even column/.append style={column sep=0.5em},
      },
    ]

    \nextgroupplot[
      title={ARM Host Setup},
      title style={font=\normalsize, yshift=-7pt},
      ymin=0,
      ymax=45,
      legend to name=grouplegend_resnet18,
      extra y ticks={1},
      extra y tick style={major tick length=0pt, grid style={red, thick}},
      extra y tick labels={}
    ]

    \addplot [
      draw=myblue,
      fill=myblue!70,
      every node near coord/.append style={text=myblue}
    ] coordinates {(25, 8.7) (50, 8.8) (100, 30.9) (200, 30.9)};

    \addplot [
      draw=mygreen,
      fill=mygreen!70,
      every node near coord/.append style={text=mygreen}
    ] coordinates {(25, 7.8) (50, 7.9) (100, 21.1) (200, 21.1)};

    \addplot [
      draw=myred,
      fill=myred!70,
      every node near coord/.append style={text=myred}
    ] coordinates {(25, 4.2) (50, 4.3) (100, 6.0) (200, 6.0)};

    \addplot [
      draw=myorange,
      fill=myorange!70,
      every node near coord/.append style={text=myorange}
    ] coordinates {(25, 2.7) (50, 2.7) (100, 3.1) (200, 3.1)};

    \pgfplotsextra{
      \pgfkeysgetvalue{/pgfplots/ymax}{\ymax}
      \pgfmathsetmacro{\yrel}{1/\ymax}
    }
    \draw[red, thick] (rel axis cs:0,\yrel) -- (rel axis cs:1,\yrel);

    \addlegendentry{\qty{50}{\nano\second}}
    \addlegendentry{\qty{100}{\nano\second}}
    \addlegendentry{\qty{500}{\nano\second}}
    \addlegendentry{\qty{1000}{\nano\second}}

    \nextgroupplot[
      title={x86 Host Setup},
      title style={font=\normalsize, yshift=-7pt},
      ymin=0,
      ymax=10,
    ]

    \addplot [
      draw=myblue,
      fill=myblue!70,
      every node near coord/.append style={text=myblue}
    ] coordinates {(25, 4.7) (50, 4.8) (100, 7.3) (200, 7.3)};

    \addplot [
      draw=mygreen,
      fill=mygreen!70,
      every node near coord/.append style={text=mygreen}
    ] coordinates {(25, 3.2) (50, 3.2) (100, 4.0) (200, 4.0)};

    \addplot [
      draw=myred,
      fill=myred!70,
      every node near coord/.append style={text=myred}
    ] coordinates {(25, 1.0) (50, 1.0) (100, 1.0) (200, 1.0)};

    \addplot [
      draw=myorange,
      fill=myorange!70,
      every node near coord/.append style={text=myorange}
    ] coordinates {(25, 1.0) (50, 1.0) (100, 1.0) (200, 1.0)};

    \pgfplotsextra{
      \pgfkeysgetvalue{/pgfplots/ymax}{\ymax}
      \pgfmathsetmacro{\yrel}{1/\ymax}
    }
    \draw[red, thick] (rel axis cs:0,\yrel) -- (rel axis cs:1,\yrel);

  \end{groupplot}

  \node at (current bounding box.south) [below=-3pt] {\pgfplotslegendfromname{grouplegend_resnet18}};
\end{tikzpicture}
  \caption{Optimal partitioning speedup for \acs{resnet}-18 across \ac{mvm} latencies and crossbar counts}
  \label{fig:model_speedup_resnet18}
\end{figure}

\cref{fig:model_speedup_resnet18} reports \acs{resnet}-18 speedups over \ac{cpu}-only execution. On ARM, the speedup reaches \(\qty{30.9}{\times}\) at \qty{50}{\nano\second} \ac{mvm} latency with \num{100} or more crossbars available. Even at \qty{100}{\nano\second} \ac{mvm} latency, the speedup remains high at \(\qty{21.1}{\times}\), demonstrating substantial acceleration potential for edge deployments. Beyond \num{100} crossbars, speedups saturate because all profitable operators are already offloaded. On x86, the peak speedup is \(\qty{7.3}{\times}\) at \qty{50}{\nano\second} \ac{mvm} latency, reflecting the stronger baseline performance of the desktop processor. For \ac{mvm} latencies above \qty{500}{\nano\second}, offloading on x86 is no longer beneficial, because the layers can be executed faster on the \ac{cpu} alone.

\begin{figure}[t]
  \centering
  \begin{tikzpicture}
  \begin{groupplot}[
      group style={
        group size=2 by 1,
        horizontal sep=2em,
        xlabels at=edge bottom,
        ylabels at=edge left,
      },
      ybar=0pt,
      width=0.55\linewidth,
      height=3.25cm,
      enlarge x limits=0.2,
      ylabel={Speedup},
      ylabel style={font=\footnotesize},
      xlabel={Number of Crossbars},
      xlabel style={font=\footnotesize, yshift=+3pt},
      symbolic x coords={25, 50, 100, 200},
      xtick=data,
      xticklabels={25, 50, 100, 200},
      tick label style={font=\footnotesize},
      x tick label style={yshift=+1pt},
      major tick length=5pt,
      tick align=center,
      tick pos=left,
      tick style={solid, black},
      ymajorgrids=true,
      major grid style={solid, gray!50},
      set layers=custom,
      /pgf/bar width=5pt,
      nodes near coords,
      nodes near coords align={vertical},
      every node near coord/.append style={
        font=\tiny,
        rotate=90,
        anchor=west,
        xshift=-2pt,
        /pgf/number format/.cd,
        fixed,
        precision=1,
        zerofill,
        /tikz/.cd,
      },
      legend style={
        at={(0.5,-0.45)},
        anchor=north,
        legend columns=-1,
        font=\footnotesize,
        draw=black,
        fill=white,
        /tikz/every even column/.append style={column sep=0.5em},
      },
    ]

    \nextgroupplot[
      title={ARM Host Setup},
      title style={font=\normalsize, yshift=-7pt},
      ymin=0,
      ymax=45,
      legend to name=grouplegend_comparison,
      extra y ticks={1},
      extra y tick style={major tick length=0pt, grid style={red, thick}},
      extra y tick labels={}
    ]

    \addplot [
      draw=myblue,
      fill=myblue!70,
      every node near coord/.append style={text=myblue}
    ] coordinates {(25, 8.7) (50, 8.8) (100, 30.9) (200, 30.9)};

    \addplot [
      draw=mygreen,
      fill=mygreen!70,
      every node near coord/.append style={text=mygreen}
    ] coordinates {(25, 2.2) (50, 6.4) (100, 14.6) (200, 14.9)};

    \addplot [
      draw=myred,
      fill=myred!70,
      every node near coord/.append style={text=myred}
    ] coordinates {(25, 2.1) (50, 3.5) (100, 3.9) (200, 3.9)};

    \addplot [
      draw=myorange,
      fill=myorange!70,
      every node near coord/.append style={text=myorange}
    ] coordinates {(25, 1.4) (50, 1.9) (100, 3.1) (200, 3.1)};

    \pgfplotsextra{
      \pgfkeysgetvalue{/pgfplots/ymax}{\ymax}
      \pgfmathsetmacro{\yrel}{1/\ymax}
    }
    \draw[red, thick] (rel axis cs:0,\yrel) -- (rel axis cs:1,\yrel);

    \addlegendentry{\acs{resnet}-18}
    \addlegendentry{\acs{resnet}-50}
    \addlegendentry{MobileNetV2}
    \addlegendentry{\acs{yolo}v5n}

    \nextgroupplot[
      title={x86 Host Setup},
      title style={font=\normalsize, yshift=-7pt},
      ymin=0,
      ymax=10,
    ]

    \addplot [
      draw=myblue,
      fill=myblue!70,
      every node near coord/.append style={text=myblue}
    ] coordinates {(25, 4.7) (50, 4.8) (100, 7.3) (200, 7.3)};

    \addplot [
      draw=mygreen,
      fill=mygreen!70,
      every node near coord/.append style={text=mygreen}
    ] coordinates {(25, 2.0) (50, 2.8) (100, 3.1) (200, 3.1)};

    \addplot [
      draw=myred,
      fill=myred!70,
      every node near coord/.append style={text=myred}
    ] coordinates {(25, 1.2) (50, 1.2) (100, 1.2) (200, 1.2)};

    \addplot [
      draw=myorange,
      fill=myorange!70,
      every node near coord/.append style={text=myorange}
    ] coordinates {(25, 1.4) (50, 1.7) (100, 1.9) (200, 1.9)};

    \pgfplotsextra{
      \pgfkeysgetvalue{/pgfplots/ymax}{\ymax}
      \pgfmathsetmacro{\yrel}{1/\ymax}
    }
    \draw[red, thick] (rel axis cs:0,\yrel) -- (rel axis cs:1,\yrel);

  \end{groupplot}

  \node at (current bounding box.south) [below=-3pt] {\pgfplotslegendfromname{grouplegend_comparison}};
\end{tikzpicture}%
  \caption{Optimal partitioning speedup at \qty{50}{\nano\second} \acs{mvm} latency across models and crossbar counts}
  \label{fig:model_speedup_comparison}
\end{figure}
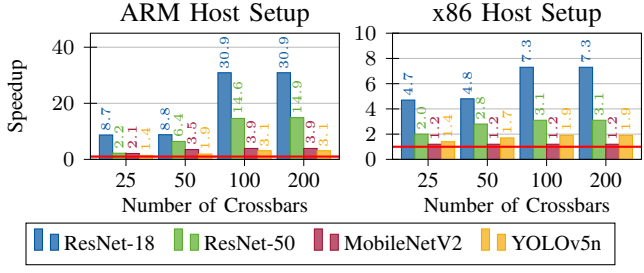

\cref{fig:model_speedup_comparison} extends the analysis to all evaluated architectures at \qty{50}{\nano\second} \ac{mvm} latency. \acs{resnet}-18 exhibits the highest speedups, followed by \acs{resnet}-50. MobileNetV2 shows limited acceleration potential because its depthwise separable convolutions cannot be efficiently mapped to \ac{rram} crossbars. Each depthwise filter operates on a single input channel, reducing the operation to a vector-vector multiplication that leaves most crossbar cells idle. \acs{yolo}v5n achieves peak speedups of \(\qty{3.1}{\times}\) on ARM and \(\qty{1.9}{\times}\) on x86. For most models, scaling beyond \num{100} crossbars yields diminishing returns. As \ac{mvm} latency increases, speedups decrease proportionally across all models.

\subsection{Operator Allocation}
\label{subsec:operator_allocation}

\begin{figure}[!b]
  \centering
  \input{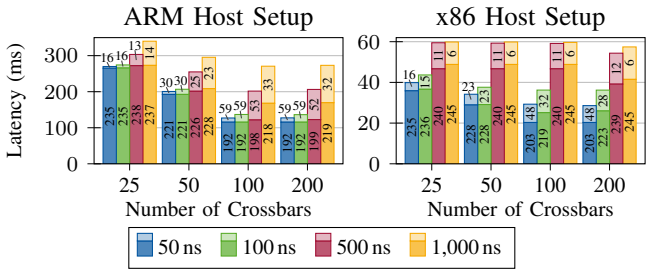}
  \caption{Latency distribution between \ac{cpu} (darker shade) and \ac{cim} accelerator (lighter shade) after optimal partitioning of \acs{yolo}v5n. Bar values indicate operator counts per device.}
  \label{fig:operator_allocation_yolov5n}
\end{figure}

\cref{fig:operator_allocation_yolov5n} shows the latency share and operator counts for \acs{yolo}v5n, which is representative of all evaluated models.

On ARM with \num{25} crossbars at \qty{50}{\nano\second} \ac{mvm} latency, only \num{16} of \num{251} operators are offloaded, as the crossbar budget cannot accommodate more layers. Scaling to \num{100} crossbars nearly quadruples the offloaded count to \num{59}, more than halving the total latency from \qty{270}{\milli\second} to \qty{127}{\milli\second}. Increasing to \num{200} crossbars does not change the allocation, confirming that crossbar scaling saturates once all compute-intensive layers are covered. A higher \ac{mvm} latency reduces offloading drastically, as it becomes less profitable. For instance, at \qty{1000}{\nano\second} \ac{mvm} latency and \num{100} crossbars, only \num{33} operators are offloaded.

On x86, the stronger host performance leads to more conservative offloading. At \qty{50}{\nano\second} \ac{mvm} latency and \num{100} crossbars, only \num{48} operators are offloaded compared to \num{59} on ARM. Furthermore, the impact of a high \ac{mvm} latency is more severe. At \qty{1000}{\nano\second} \ac{mvm} latency, offloading reduces to \num{6} operators for all crossbar counts.

This confirms that the \ac{ilp} adapts to platform and hardware parameters by balancing computation and communication costs.

\subsection{Scalability of the \acs{ilp} Approach}
\label{subsec:ilp_scalability}

\begin{table}[t]
  \centering
  \caption{\acs{ilp} formulation statistics for workload partitioning on the ARM setup}
  \label{tab:ilp_stats}
  \begin{tabular}{@{\hspace{0.5em}} c @{\hspace{0.5em}} S[table-format=3] @{\hspace{0.5em}} S[table-format=4] @{\hspace{0.5em}} S[table-format=4] @{\hspace{0.5em}} c @{\hspace{0.5em}} S[table-format=1.2] @{\hspace{0.5em}}}
    \toprule
    Model & {\makecell{Operators}} & {\makecell{Decision\\Variables}} & {\makecell{Constraints}} & \makecell{Mean Solving\\Time (\unit{\second})} & {\makecell{Mean \acs{mip}\\Gap (\unit{\percent})}} \\
    \midrule
    \acs{resnet}-18 & 38 & 44 & 204 & \(<\)\num{0.01} & 0.02 \\
    \acs{resnet}-50 & 79 & 91 & 412 & \hspace{0.75em}\num{0.24} & 0.37 \\
    MobileNetV2 & 71 & 71 & 331 & \(<\)\num{0.01} & 0.07 \\
    \acs{yolo}v5n & 251 & 3041 & 7035 & \num{503} & 6.97 \\
    \bottomrule
  \end{tabular}
\end{table}

\cref{tab:ilp_stats} summarizes the \ac{ilp} sizes and solver metrics on ARM. The \acp{ilp} of \acs{resnet}-18 and MobileNetV2 are solved in under \qty{0.01}{\second} with a mean \ac{mip} gap below \qty{0.1}{\percent}. The larger \acs{resnet}-50 converges in \qty{0.24}{\second} with a mean \ac{mip} gap of \qty{0.37}{\percent}. This confirms that larger linear networks remain manageable.

\acs{yolo}v5n exhibits a substantial increase in problem complexity due to its heavily branched architecture. Despite having only \(\qty{3.2}{\times}\) more operators than \acs{resnet}-50, its branching topology leads to \(\qty{33}{\times}\) more decision variables and \(\qty{17}{\times}\) more constraints. Both arise from the complex execution orderings across parallel paths. The solver often reaches the \qty{600}{\second} timeout, yielding solutions with a mean \ac{mip} gap of \qty{6.97}{\percent}. As discussed in \cref{subsec:ilp}, the \ac{lp} relaxation bound may be significantly looser than the true \ac{ilp} optimum, so these solutions may still be optimal or near-optimal.

These results show that graph branching, rather than operator count, drives \ac{ilp} complexity and solving time. 

\section{Conclusion}
\label{sec:conclusion}

We presented an \ac{ilp}-based \ac{ml} workload partitioning framework for heterogeneous systems comprising a host \ac{cpu} and an \ac{rram}-based \ac{cim} accelerator. It combines empirical host \ac{cpu} profiling with analytical \ac{cim} and bus latency models. The \ac{ilp} formulation minimizes the end-to-end inference latency under \ac{rram} memory budget and operator support constraints, exploiting both intra-accelerator and inter-device parallelism.

Our experiments on \acs{resnet}-18, \acs{resnet}-50, MobileNetV2, and \acs{yolo}v5n demonstrate speedups of up to \(\qty{30.9}{\times}\) over an ARM edge \ac{cpu} and \(\qty{7.3}{\times}\) over a high-performance x86 \ac{cpu}. The optimal strategy consistently offloads compute-intensive convolutions. The \ac{dse} shows that the \ac{mvm} latency dominates hardware sensitivity, while crossbar scaling saturates once all compute-intensive layers are covered. The \ac{ilp} formulation scales well for linear and sparsely branched architectures and remains practical for all evaluated edge-relevant models. Future work could explore heuristic or hybrid methods to reduce solving time for complex branching topologies.

\bibliographystyle{IEEEtran}
\balance
\bibliography{bibtexentry}

\begin{thebibliography}{10}
\providecommand{\url}[1]{#1}
\csname url@samestyle\endcsname
\providecommand{\newblock}{\relax}
\providecommand{\bibinfo}[2]{#2}
\providecommand{\BIBentrySTDinterwordspacing}{\spaceskip=0pt\relax}
\providecommand{\BIBentryALTinterwordstretchfactor}{4}
\providecommand{\BIBentryALTinterwordspacing}{\spaceskip=\fontdimen2\font plus
\BIBentryALTinterwordstretchfactor\fontdimen3\font minus \fontdimen4\font\relax}
\providecommand{\BIBforeignlanguage}[2]{{%
\expandafter\ifx\csname l@#1\endcsname\relax
\typeout{** WARNING: IEEEtran.bst: No hyphenation pattern has been}%
\typeout{** loaded for the language `#1'. Using the pattern for}%
\typeout{** the default language instead.}%
\else
\language=\csname l@#1\endcsname
\fi
#2}}
\providecommand{\BIBdecl}{\relax}
\BIBdecl

\bibitem{amrouch2021ReliableInMemoryComputing}
H.~Amrouch, N.~Du, A.~Gebregiorgis, S.~Hamdioui, and I.~Polian, ``Towards {{Reliable In-Memory Computing}}:{{From Emerging Devices}} to {{Post-von-Neumann Architectures}},'' in \emph{2021 {{IFIP}}/{{IEEE}} 29th {{International Conference}} on {{Very Large Scale Integration}} ({{VLSI-SoC}})}, Oct. 2021, pp. 1--6.

\bibitem{shafiee2016ISAACConvolutionalNeural}
A.~Shafiee, A.~Nag, N.~Muralimanohar, R.~Balasubramonian, J.~P. Strachan, M.~Hu, R.~S. Williams, and V.~Srikumar, ``{{ISAAC}}: A convolutional neural network accelerator with in-situ analog arithmetic in crossbars,'' \emph{SIGARCH Comput. Archit. News}, vol.~44, no.~3, pp. 14--26, Jun. 2016.

\bibitem{chi2016PRIMENovelProcessinginMemory}
P.~Chi, S.~Li, C.~Xu, T.~Zhang, J.~Zhao, Y.~Liu, Y.~Wang, and Y.~Xie, ``{{PRIME}}: {{A Novel Processing-in-Memory Architecture}} for {{Neural Network Computation}} in {{ReRAM-Based Main Memory}},'' in \emph{2016 {{ACM}}/{{IEEE}} 43rd {{Annual International Symposium}} on {{Computer Architecture}} ({{ISCA}})}, Jun. 2016, pp. 27--39.

\bibitem{ankit2019PUMAProgrammableUltraefficient}
A.~Ankit, I.~E. Hajj, S.~R. Chalamalasetti, G.~Ndu, M.~Foltin, R.~S. Williams, P.~Faraboschi, W.-m.~W. Hwu, J.~P. Strachan, K.~Roy, and D.~S. Milojicic, ``{{PUMA}}: {{A Programmable Ultra-efficient Memristor-based Accelerator}} for {{Machine Learning Inference}},'' in \emph{Proceedings of the {{Twenty-Fourth International Conference}} on {{Architectural Support}} for {{Programming Languages}} and {{Operating Systems}}}, ser. {{ASPLOS}} '19.\hskip 1em plus 0.5em minus 0.4em\relax New York, NY, USA: Association for Computing Machinery, Apr. 2019, pp. 715--731.

\bibitem{wan2022ComputeinmemoryChipBased}
W.~Wan, R.~Kubendran, C.~Schaefer, S.~B. Eryilmaz, W.~Zhang, D.~Wu, S.~Deiss, P.~Raina, H.~Qian, B.~Gao, S.~Joshi, H.~Wu, H.-S.~P. Wong, and G.~Cauwenberghs, ``A compute-in-memory chip based on resistive random-access memory,'' \emph{Nature}, vol. 608, no. 7923, pp. 504--512, Aug. 2022.

\bibitem{zahoor2020ResistiveRandomAccess}
F.~Zahoor, T.~Z. Azni~Zulkifli, and F.~A. Khanday, ``Resistive {{Random Access Memory}} ({{RRAM}}): An {{Overview}} of {{Materials}}, {{Switching Mechanism}}, {{Performance}}, {{Multilevel Cell}} (mlc) {{Storage}}, {{Modeling}}, and {{Applications}},'' \emph{Nanoscale Research Letters}, vol.~15, no.~1, pp. 90--115, Apr. 2020.

\bibitem{gao2024StaticSchedulingWeight}
X.~Gao, H.~Wang, Y.~Chen, Y.~Zhang, Z.~Shen, and L.~Ju, ``Static {{Scheduling}} of {{Weight Programming}} for {{DNN Acceleration}} with {{Resource Constrained PIM}},'' \emph{ACM Trans. Embed. Comput. Syst.}, vol.~23, no.~6, pp. 89:1--89:22, Sep. 2024.

\bibitem{he2016DeepResidualLearning}
K.~He, X.~Zhang, S.~Ren, and J.~Sun, ``Deep {{Residual Learning}} for {{Image Recognition}},'' in \emph{2016 {{IEEE Conference}} on {{Computer Vision}} and {{Pattern Recognition}} ({{CVPR}})}.\hskip 1em plus 0.5em minus 0.4em\relax Las Vegas, NV, USA: {Institute of Electrical and Electronics Engineers (IEEE)}, Jun. 2016, pp. 770--778.

\bibitem{sandler2018MobileNetV2InvertedResiduals}
M.~Sandler, A.~Howard, M.~Zhu, A.~Zhmoginov, and L.-C. Chen, ``{{MobileNetV2}}: {{Inverted Residuals}} and {{Linear Bottlenecks}},'' in \emph{2018 {{IEEE}}/{{CVF Conference}} on {{Computer Vision}} and {{Pattern Recognition}}}.\hskip 1em plus 0.5em minus 0.4em\relax Salt Lake City, UT, USA: {Institute of Electrical and Electronics Engineers (IEEE)}, Jun. 2018, pp. 4510--4520.

\bibitem{jocher2020UltralyticsYOLOv5}
G.~Jocher, ``Ultralytics {{YOLOv5}},'' Zenodo, 2020.

\bibitem{park2025COMPASSCompilerFramework}
J.~Park, J.~Choe, D.~Kim, and J.-J. Kim, ``{{COMPASS}}: {{A Compiler Framework}} for {{Resource-Constrained Crossbar-Array Based In-Memory Deep Learning Accelerators}},'' in \emph{2025 {{Design}}, {{Automation}} \& {{Test}} in {{Europe Conference}} ({{DATE}})}, Mar. 2025, pp. 1--7.

\bibitem{swaidan2019RRAMEnduranceRetention}
Z.~Swaidan, R.~Kanj, J.~El~Hajj, E.~Saad, and F.~Kurdahi, ``{{RRAM Endurance}} and {{Retention}}: {{Challenges}}, {{Opportunities}} and {{Implications}} on {{Reliable Design}},'' in \emph{2019 26th {{IEEE International Conference}} on {{Electronics}}, {{Circuits}} and {{Systems}} ({{ICECS}})}, Nov. 2019, pp. 402--405.

\bibitem{kang2017NeurosurgeonCollaborativeIntelligence}
Y.~Kang, J.~Hauswald, C.~Gao, A.~Rovinski, T.~Mudge, J.~Mars, and L.~Tang, ``Neurosurgeon: {{Collaborative Intelligence Between}} the {{Cloud}} and {{Mobile Edge}},'' in \emph{Proceedings of the {{Twenty-Second International Conference}} on {{Architectural Support}} for {{Programming Languages}} and {{Operating Systems}}}, ser. {{ASPLOS}} '17.\hskip 1em plus 0.5em minus 0.4em\relax New York, NY, USA: Association for Computing Machinery, Apr. 2017, pp. 615--629.

\bibitem{viramontes2023NeuralNetworkPartitioning}
R.~Viramontes and A.~Davoodi, ``Neural {{Network Partitioning}} for {{Fast Distributed Inference}},'' in \emph{2023 24th {{International Symposium}} on {{Quality Electronic Design}} ({{ISQED}})}, Apr. 2023, pp. 1--7.

\bibitem{zeng2021CoEdgeCooperativeDNN}
L.~Zeng, X.~Chen, Z.~Zhou, L.~Yang, and J.~Zhang, ``{{CoEdge}}: {{Cooperative DNN Inference With Adaptive Workload Partitioning Over Heterogeneous Edge Devices}},'' \emph{IEEE/ACM Transactions on Networking}, vol.~29, no.~2, pp. 595--608, Apr. 2021.

\bibitem{taufique2025HiDPHierarchicalDNN}
Z.~Taufique, A.~Vyas, A.~Miele, P.~Liljeberg, and A.~Kanduri, ``{{HiDP}}:{{Hierarchical DNN Partitioning}} for {{Distributed Inference}} on {{Heterogeneous Edge Platforms}},'' in \emph{2025 {{Design}}, {{Automation}} \& {{Test}} in {{Europe Conference}} ({{DATE}})}, Mar. 2025, pp. 1--7.

\bibitem{li2023MathematicalFrameworkOptimizing}
W.~Li, Y.~Han, and X.~Chen, ``Mathematical {{Framework}} for {{Optimizing Crossbar Allocation}} for {{ReRAM-based CNN Accelerators}},'' \emph{ACM Trans. Des. Autom. Electron. Syst.}, vol.~29, no.~1, pp. 21:1--21:24, Dec. 2023.

\bibitem{qu2024CIMMLCMultilevelCompilation}
S.~Qu, S.~Zhao, B.~Li, Y.~He, X.~Cai, L.~Zhang, and Y.~Wang, ``{{CIM-MLC}}: {{A Multi-level Compilation Stack}} for {{Computing-In-Memory Accelerators}},'' in \emph{Proceedings of the 29th {{ACM International Conference}} on {{Architectural Support}} for {{Programming Languages}} and {{Operating Systems}}, {{Volume}} 2}, ser. {{ASPLOS}} '24, vol.~2.\hskip 1em plus 0.5em minus 0.4em\relax New York, NY, USA: Association for Computing Machinery, Apr. 2024, pp. 185--200.

\bibitem{gurobioptimizationllc2026GurobiOptimizerReference}
\BIBentryALTinterwordspacing
{Gurobi Optimization, Llc}, ``Gurobi {{Optimizer Reference Manual}},'' 2026. [Online]. Available: \url{https://www.gurobi.com}
\BIBentrySTDinterwordspacing

\bibitem{onnxdevelopers2026ONNXOpenNeural}
\BIBentryALTinterwordspacing
{Onnx Developers}, ``{{ONNX}}: {{Open Neural Network Exchange}},'' 2026. [Online]. Available: \url{https://onnx.ai/}
\BIBentrySTDinterwordspacing

\bibitem{onnxruntimedevelopers2021ONNXRuntime}
\BIBentryALTinterwordspacing
{Onnx Runtime Developers}, ``{{ONNX Runtime}},'' 2021. [Online]. Available: \url{https://onnxruntime.ai/}
\BIBentrySTDinterwordspacing

\bibitem{pelke2023MappingCNNsMulticore}
R.~Pelke, N.~Bosbach, J.~Cubero, F.~Staudigl, R.~Leupers, and J.~M. Joseph, ``Mapping of {{CNNs}} on multi-core {{RRAM-based CIM}} architectures,'' in \emph{2023 {{IFIP}}/{{IEEE}} 31st {{International Conference}} on {{Very Large Scale Integration}} ({{VLSI-SoC}})}, Oct. 2023, pp. 1--6.

\bibitem{pelke2026MixedPrecisionTrainingCompilation}
R.~Pelke, J.~Klein, J.~{Cubero-Cascante}, N.~Bosbach, J.~M. Joseph, and R.~Leupers, ``Mixed-{{Precision Training}} and {{Compilation}} for {{RRAM-based Computing-in-Memory Accelerators}},'' in \emph{2026 {{Design}}, {{Automation}} \& {{Test}} in {{Europe Conference}} ({{DATE}})}.\hskip 1em plus 0.5em minus 0.4em\relax Verona, Italy: IEEE, Apr. 2026, pp. 1--7.

\bibitem{jeon2025LowRankCompressionIMC}
K.~E. Jeon, J.~Rhe, and J.~H. Ko, ``Low-{{Rank Compression}} for {{IMC Arrays}},'' in \emph{2025 {{Design}}, {{Automation}} \& {{Test}} in {{Europe Conference}} ({{DATE}})}.\hskip 1em plus 0.5em minus 0.4em\relax Lyon, France: IEEE, Mar. 2025, pp. 1--7.

\bibitem{jacob2018QuantizationTrainingNeural}
B.~Jacob, S.~Kligys, B.~Chen, M.~Zhu, M.~Tang, A.~Howard, H.~Adam, and D.~Kalenichenko, ``Quantization and {{Training}} of {{Neural Networks}} for {{Efficient Integer-Arithmetic-Only Inference}},'' in \emph{2018 {{IEEE}}/{{CVF Conference}} on {{Computer Vision}} and {{Pattern Recognition}}}, Jun. 2018, pp. 2704--2713.

\end{thebibliography}

\end{document}